\documentclass[fleqn,10pt]{article}
\usepackage[utf8]{inputenc}
\usepackage[T1]{fontenc}
\usepackage{graphicx}
\usepackage{amsmath}
\usepackage{xcolor}
\usepackage{float}
\usepackage{anysize}
\usepackage{pdflscape}
\usepackage{authblk}
\usepackage{hyperref}
\marginsize{2cm}{2cm}{2cm}{2cm}
\newcommand{\opt}{\text{opt}}

\title{Spatially multiplexed single-photon sources based on binary-tree multiplexers with optimized structure}
\author[1,2]{Matyas Mechler}
\author[1,2,*]{Peter Adam}
\affil[1]{Institute for Solid State Physics and Optics, Wigner Research Centre for Physics, P.O. Box 49, H-1525 Budapest, Hungary}
\affil[2]{Institute of Physics, University of P\'ecs, Ifj\'us\'ag \'utja 6, H-7624 P\'ecs, Hungary}
\affil[*]{Corresponding author: adam.peter@wigner.hu}
\date{}
\begin{document}
\flushbottom
\maketitle
\begin{abstract}
We develop a method for optimizing the structure of general binary-tree multiplexers realized with asymmetric photon routers aiming at improving the performance of spatially multiplexed single-photon sources.
Our procedure systematically considers all possible binary-tree multiplexers that can be formed by a certain number of photon routers.
Using this method one can select the multiplexer structure that leads to the highest single-photon probability for a given set of loss parameters characterizing the system.
We determine the optimal general binary-tree multiplexers for experimentally realizable values of the transmission coefficients of the photon routers and that of the detector efficiency.
We show that single-photon sources based on such optimal multiplexers yield higher single-photon probabilities than that can be achieved with single-photon sources based on any other spatial multiplexer considered in the literature.
Our approach improves the performance of multiplexed single-photon sources even for small system sizes which is the typical situation in current experiments.
\end{abstract}

\section{Introduction}
The substantial role of single-photon sources (SPSs) in the effective realization of a number of experiments in the fields of quantum information processing and photonic quantum technology keeps their development in the focus of research~\cite{EisamanRSI2011, MScott2020}.
Multiplexed SPSs can be promising candidates for yielding indistinguishable single photons in near-perfect spatial modes with known polarization on demand.
Such sources are based on heralded SPSs\cite{PittmanOC2005, Mosley2008, Brida2011, RamelowOE2013, MassaroNJP2019, LuganiOE2020} in which the detection of one member of a correlated photon pair generated in nonlinear optical processes heralds the presence of its twin photon.
In heralded SPSs the multiphoton noise originating from the inherent probabilistic nature of the nonlinear processes can be reduced by using single-photon detectors with photon number resolving capabilities for heralding, or by decreasing the mean photon number of the generated photon pairs.
Multiplexing several sources of heralded photons can compensate for the decrease of the probability of successful heralding caused by the reduction of the mean photon number.
Multiplexing can be realized by suitable switching systems in which heralded photons generated in particular multiplexed units are rerouted to a single output mode.
Various schemes have been proposed for single-photon sources based on spatial \cite{Migdall2002, ShapiroWong2007, Ma2010, Collins2013, MazzarellaPRA2013, Meany2014, Adam2014, BonneauNJP2015, Francis2016, KiyoharaOE2016, Bodog2016, AdamOE2022} and temporal multiplexing \cite{Pittman2002, Jeffrey2004, Mower2011, Adam2014, Schmiegelow2014, Kaneda2015, Rohde2015, XiongNC2016, Hoggarth2017, HeuckNJP2018, Kaneda2019, Lee2019, MagnoniQIP2019}, and some of them have been successfully implemented in experiments \cite{Ma2010, Collins2013, Meany2014, Kaneda2015, Francis2016, KiyoharaOE2016, XiongNC2016, Hoggarth2017, Kaneda2019}.

In real multiplexed SPSs, the presence of various losses leads to the degradation of the performance \cite{MazzarellaPRA2013, BonneauNJP2015}.
The output single-photon probability of these systems can be maximized by determining the optimal number of multiplexed units and the mean number of photon pairs generated in the units.
The optimization can be performed by applying the full statistical theories developed for the description of such systems \cite{Adam2014, Bodog2016, Bodog2020, AdamPRA2022, AdamOE2022}.
According to the analyses, state-of-the-art multiplexed SPSs realized with low-loss optical elements can yield high achievable single-photon probabilities with low multiphoton contribution\cite{Bodog2020, AdamPRA2022, AdamOE2022, AdamOE2023}.

In spatially multiplexed single-photon sources, several individual pulsed heralded photon sources are applied in parallel.
These sources can be realized by using physically separate nonlinear processes or in separate spatial modes of a single process.
After a successful heralding event in one of the heralded sources, a spatial multiplexer composed of a set of binary photon routers is used to reroute the corresponding heralded signal photons to a single output.
Main types of spatial multiplexers considered thus far in the literature are symmetric (complete binary-tree), asymmetric (chain-like), and incomplete binary-tree multiplexers.
Successful experimental realizations of multiplexed SPSs based on symmetric multiplexers have been reported up to four multiplexed units by using spontaneous parametric down-conversion in bulk crystals \cite{Ma2010,KiyoharaOE2016} and waveguides \cite{Meany2014}, and by using spontaneous four-wave mixing up to two multiplexed units in photonic crystal fibers \cite{Collins2013, Francis2016}.
Theoretical analyses showed that a particular multiplexer structure can outperform the other for a certain range of the loss parameters when applied in SPSs\cite{AdamOE2022, AdamOE2023}.
Hence, finding novel multiplexing schemes that can further improve the performance of SPSs is an important goal of the researches on multiplexed SPSs.

In the present paper, we consider general binary-tree multiplexers, that is, all binary-tree multiplexers that can be formed by a given number of photon routers.
We develop a systematic method for finding the binary-tree structure that leads to a single-photon source with the highest performance for a given set of loss parameters.
We analyze the performance of SPSs based on optimal general binary-tree multiplexers in detail.

\section{Single-photon sources based on general binary-tree multiplexers}\label{sec:2}
A single-photon source based on a general spatial multiplexer contains a set of multiplexed units (MUs) and a multiplexer, that is, a multiport routing device.
The MUs are heralded single-photon sources independent of each other.
Each multiplexed unit contains a nonlinear photon pair source and a detector for detecting the idler photons of the photon pairs.
Detection events in the multiplexed units herald the presence of the corresponding signal photons which in turn are directed to a single output by the multiplexer.
In this paper, we consider general spatial multiplexers built of binary photon routers (PRs), that is, routing elements with two inputs and a single output.
The output of a PR can be connected to any of the inputs of another PR.
Several building logics were hitherto analyzed in the literature such as symmetric multiplexers also referred to as complete binary-tree multiplexers, asymmetric (chain-like) structures, and various incomplete binary-tree multiplexers constructed by following either a geometric logic or a transmission-based logic.
Now we do not pose any restrictions on the structure the PRs are connected to each other, that is, we consider all possible binary trees that can be constructed by using a certain number of PRs.
Consequently, we introduce the term \emph{general binary-tree multiplexer} (GBM) to refer to these multiplexers.
PRs are generally asymmetric, that is, the photon losses characterizing the two input ports of the PR are different.
An asymmetric PR can be realized easily by using bulk-optical elements~\cite{Bodog2016,AdamOE2022}.
In the cited papers, the terms \emph{transmission} and \emph{reflection efficiencies} were used for the efficiencies characterizing the two input ports of the PR with the corresponding notations $V_t$ and $V_r$, respectively.
In the present paper, we keep these notations for the inputs of the PRs and use the term \emph{transmission coefficients} to refer to both of them.
We note that periodicity of the single-photon output is a requirement posed by most applications that can be ensured by pulsed pumping of the source generating the photon pairs.
Also, beside multiplexing, suppressing multiphoton noise in multiplexed single-photon sources can be guaranteed by applying single-photon detectors with photon-number-resolving capabilities in the MUs\cite{Divochiy2008, Lita2008, Fukuda2011, Jahanmirinejad2012, CahallOptica2017, Schmidt2018, Fukuda2019}, therefore we assume such detectors in our calculations.
We also mention that the MUs can contain an optional delay line placed into the path of the signal photon that is responsible for introducing a sufficiently long delay into the traveling time of the photon before it enters the multiplexer. This delay enables the operation of the logic controlling the routers.

\begin{figure}[tb]
\centering
\includegraphics[width=0.8\textwidth]{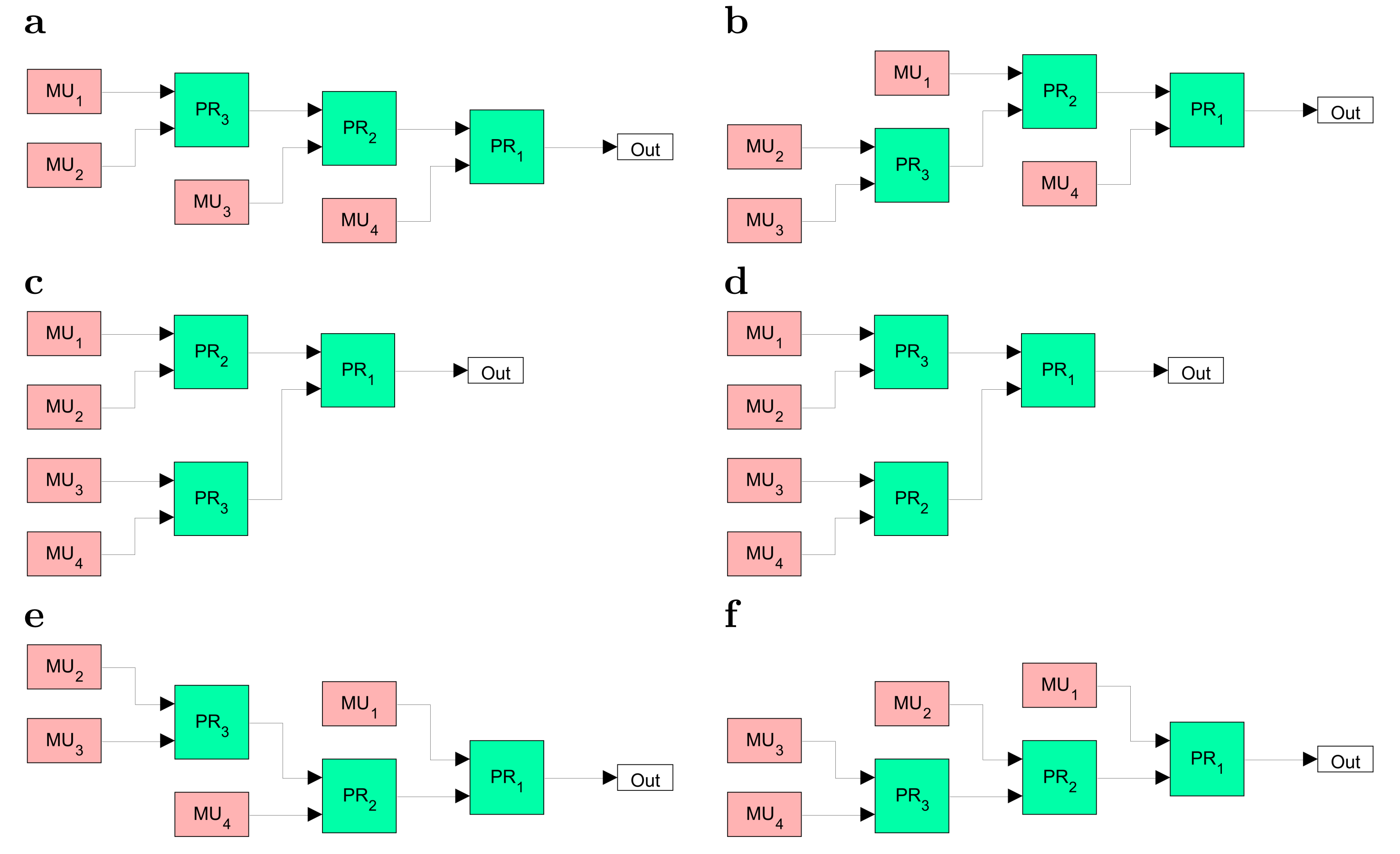}
\caption{All binary-tree multiplexers constructed by using three binary PRs. The corresponding integer sequences identifying the particular multiplexers are: (\textbf{a}) [1,1,1], (\textbf{b}) [1,1,2], (\textbf{c}) [1,1,3], (\textbf{d}) [1,2,1], (\textbf{e}) [1,2,2], (\textbf{f}) [1,2,3]. \label{fig:ThreeExample}}
\end{figure}

Our aim is to find the GBM composed of $N_R$ identical asymmetric PRs that gives the highest output single-photon probability $P_1$.
Therefore we need to test all possible different binary-tree structures.
We assume that the positioning of the PRs in a binary-tree is fixed, that is, the inputs of all routers characterized by given transmission efficiencies are in the same geometric position in the tree.
We prescribe that the numbering of the inputs of any of the multiplexers follows the geometric rule applied at the inputs of the first router.
Then we can represent a binary-tree multiplexer comprising $N_R$ PRs by a sequence of integer numbers of length $N_R$ according to the following logic.
The first number is always 1 showing that the first router is connected to the single output of the multiplexer.
The second number identifies the connection point of the second router to the first one, therefore it can take the values 1 or 2.
The $n$th number is the connection point of the $n$th PR and, accordingly, it can take any integer value between 1 and $n$.
Obviously, the number of such sequences for binary-tree multiplexers formed by $N_R$ routers is $N_R!$.
As an example, in Fig.~\ref{fig:ThreeExample} we show all the possible multiplexers comprising three PRs, and in the caption of the Figure we specify the integer sequences identifying the various multiplexers.
These sequences representing the six multiplexers in the Figure are [1,1,1], [1,1,2], [1,1,3], [1,2,1], [1,2,2], [1,2,3].
However, as we assume identical routers, the sequences [1,1,3] and [1,2,1] represent the same two-level complete binary-tree multiplexers.
To avoid this problem, we apply the following rule: we accept only sequences containing 0 or 1 increment or arbitrary decrement between subsequent elements.
This way, the sequence [1,1,3] is excluded from the above list of sequences.
It can be shown that, following this rule, those sequences representing  binary-tree multiplexers having structures identical with an already observed one can be excluded from the list of sequences for any number of PRs.
The number of sequences of length $N_R$ generated in this way provides the number of different binary-tree multiplexers $K_{N_R}$ formed by $N_R$ PRs.
It can be found that $K_{N_R}$ can be calculated as
\begin{equation}
K_{N_R}=\prod_{k=2}^{N_R}\frac{N_R+k}{k} \quad \text{for }N_R\geq 2.
\label{eq:Catalan}
\end{equation}
We note that this number is known as the $N_R$th Catalan number \cite{StanleyBookCatalan2015}.

As it was explained above, a two-port routing device, that is, a binary PR can be characterized by the transmission coefficients $V_t$ and $V_r$.
Similarly, a multiplexer which is a multiport routing device can be characterized by the \emph{total transmission coefficients} $V_n$ describing the transmission probabilities between each input and the output of the multiplexer.
These total transmission coefficients are in fact products of the transmission coefficients of the PRs, hence they can be written in the symbolic form of 
\begin{equation}
V_n=V_bV_r^jV_t^k\qquad (0\leq j,k\leq N).
\end{equation}
Here, the multiplicative factor $V_b$ termed as \emph{general transmission coefficient} characterizes all other losses experienced by the photons while propagating to the input of the multiplexer after their heralding.
A multiplexer formed by $N_R$ PRs have $N_R+1$ inputs and, accordingly, it can be characterized by $N=N_R+1$ total transmission coefficients.
In the following $N$ denotes the number of MUs that are connected to the given multiplexer.
In the next section, we will compare our results to the performance of SPSs based on asymmetric (ASYM) multiplexers.
Such multiplexers have a chain-like structure characterized by the total transmission coefficients
\begin{equation}
\begin{alignedat}{3}
& V_n = V_bV_1V_2^{n-1} \quad && \text{if} \quad && n<N, \\ 
& V_n = V_bV_2^{n-1} \quad &&  \text{if} \quad &&  n=N,\label{eq:ASM}
\end{alignedat}
\end{equation}
where $V_1$ and $V_2$ are the smaller and larger, respectively, of the transmission coefficients $V_r$ and $V_t$.

\begin{figure}[tb]
\centering
\includegraphics[width=0.8\textwidth]{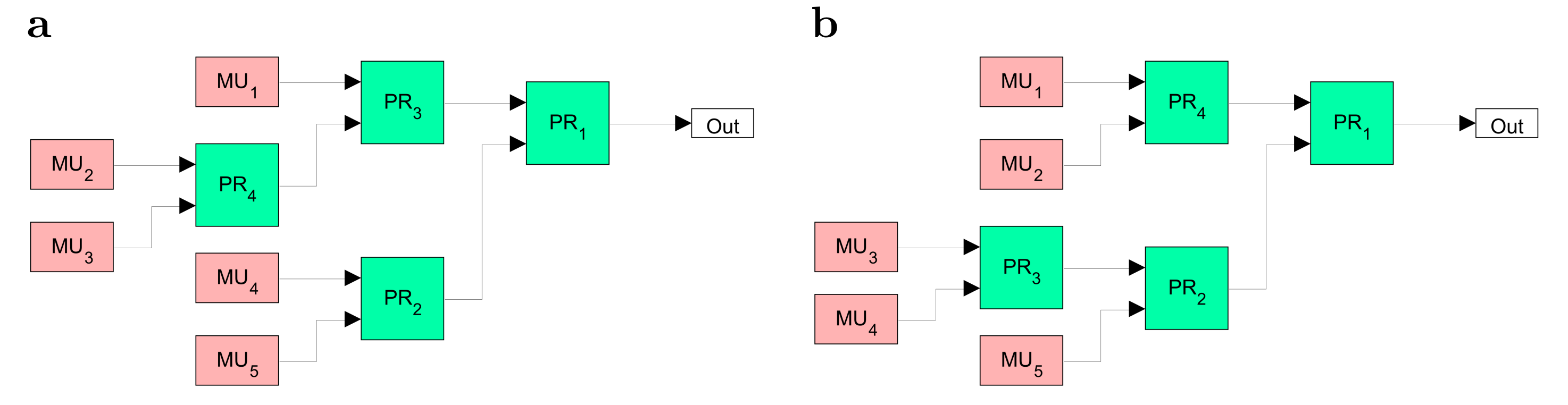}
\caption{Two binary-tree multiplexers constructed by using four binary PRs having the same sets $\{V_n\}$ of total transmission coefficients $V_n$. The corresponding integer sequences identifying the particular multiplexers are: (\textbf{a}) [1,2,1,2], (\textbf{b}) [1,2,2,1]. Assuming that the transmission coefficients of the upper and lower inputs of the PRs are denoted by $V_t$ and $V_r$, respectively, the corresponding sets are
(\textbf{a}) $\{V_t^2,V_t^2V_r,V_tV_r^2,V_tV_r,V_r^2\}$ and
(\textbf{b}) $\{V_t^2,V_tV_r,V_t^2V_r,V_tV_r^2,V_r^2\}$.
\label{fig:FourExample}}
\end{figure}
For analyzing SPSs based on GBMs, we will apply the general statistical theory developed previously for treating SPSs based on either spatial or temporal multiplexing equipped with photon-number-resolving detectors realizing any detection strategy~\cite{Bodog2020, AdamOE2022}.
We will consider only ranges of the loss parameters for which single-photon detection is certainly the optimal detection strategy.
In this case, the probability $P_i$ of obtaining $i$ photons at the output of multiplexed SPSs can be written as
\begin{eqnarray}
P_i=\left(1-P^{(D)}_1\right)^N\delta_{i,0}+\sum_{n=1}^N\left[\left(1-P^{(D)}_1\right)^{n-1}\times\sum_{l=i}^\infty P^{(D)}(1|l)P^{(\lambda)}(l)V_n(i|l)\right].\label{general_formula_var}
\end{eqnarray}
Here, the variable $l$ denotes the number of photon pairs generated by the nonlinear source in the $n$th multiplexed unit MU$_n$.
$P^{(\lambda)}(l)$ is the probability of generating $l$ pairs in a MU assuming that the mean photon number of the generated pairs, that is, the \emph{input mean photon number} is $\lambda$.
We assume that a single-mode nonlinear process with strong spectral filtering is used in the scheme.
In this case, the multiplexed SPSs can yield highly indistinguishable single photons that are required in many experiments and applications \cite{Avenhaus2008, Collins2013, KiyoharaOE2016}, and
\begin{equation}
P^{(\lambda)}(l)=\frac{\lambda^l}{(1+\lambda)^{1+l}},
\end{equation}
that is, the probability distribution of the input mean photon number is thermal.
$P^{(D)}(1|l)$ denotes the conditional probability of registering a single photon provided that $l$ photons arrive at the detector with detector efficiency $V_D$. It can be expressed as
\begin{equation}
    P^{(D)}(1|l)=lV_D(1-V_D)^{l-1}.
\end{equation}
The total probability $P^{(D)}_1$, that is, the probability of the event that a single photon is detected can be derived as
\begin{eqnarray}
P^{(D)}_1=\sum_{l=1}^\infty P^{(D)}(1|l)P^{(\lambda)}(l)=\frac{V_D\lambda}{(V_D\lambda+1)^2}.\label{eq:tpPD}
\end{eqnarray}
In our calculations, the probabilities $P^{(D)}_1$, $P^{(D)}(1|l)$, $P^{(\lambda)}(l)$, and the input mean photon number $\lambda$ are assumed to be independent of the sequential number $n$ of the multiplexed unit.

Finally, $V_n(i|l)$ is the conditional probability
of the event that $i$ photons reach the output of the multiplexer provided that $l$ signal photons arrive from the $n$th multiplexed unit into the system. This probability is expressed as
\begin{equation}
    V_n(i|l)=\binom{l}{i}V_n^i(1-V_n)^{l-i},
\end{equation}
where the total transmission coefficient $V_n$ 
characterizes the losses of the $n$th arm of the particular multiplexer. 

\begin{figure}[tb]
    \centering
    \includegraphics[width=\textwidth]{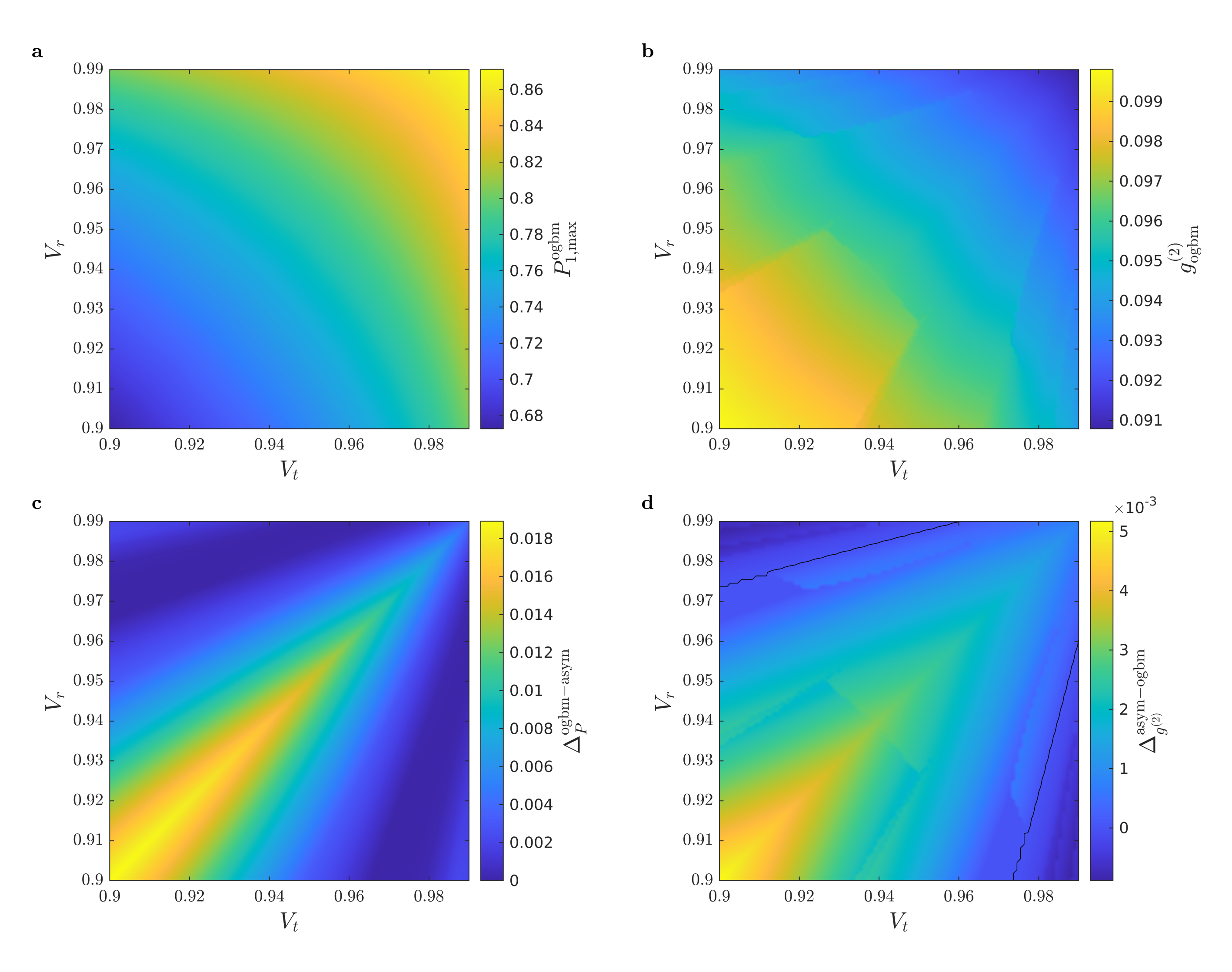}
    \caption{(\textbf{a}) The maximal single-photon probability $P_{1,\max}^{\rm ogbm}$ and (\textbf{b}) the normalized second-order autocorrelation function $g^{(2)}_{\rm ogbm}$ for SPSs based on OGBMs as functions of the transmission coefficients $V_t$ and $V_r$. (\textbf{c}) The difference $\Delta_P^{\rm ogbm - asym}$ between the maximal single-photon probabilities, and (\textbf{d}) the difference $\Delta_{g^{(2)}}^{\rm asym - ogbm}$ between the normalized second-order autocorrelation functions for SPSs based on OGBM and ASYM multiplexers, respectively, as functions of the transmission coefficients $V_t$ and $V_r$. Here the detector efficiency $V_D=0.95$ and the number of multiplexed units $N=11$. \label{fig:P1g2DP1Dg2}}
\end{figure}
From the second term of equation~\eqref{general_formula_var}, one can see that this theory assumes a priority logic controlling the multiplexed SPS that prefers the MU with the smallest sequential number $n$ if multiple heralding events happen in different MUs.
It seems plausible that by assigning smaller sequential numbers $n$ to arms with higher total transmission coefficients $V_n$, the achievable single-photon probability $P_1$ can be higher.
Consequently, it is reasonable to choose a numbering for the MUs for which the associated total transmission coefficients $V_n$ are arranged into a decreasing order, that is, $V_1\ge V_2\ge\dots\ge V_N$.
Obviously, the numbering of the multiplexer arms having identical total transmission coefficients is arbitrary.

Knowing the probabilities $P_i$ from equation~\eqref{general_formula_var}, the normalized second-order autocorrelation function can be obtained as
\begin{equation}
g^{(2)}(t=0)=\frac{\displaystyle\sum_{i=2}^\infty P_i i (i-1)}{\left(\displaystyle\sum_{i=1}^{\infty} P_i  i\right)^2}.
\end{equation}
This function quantifies the contribution of multiphoton components in the output state compared to that of the single-photon component.
In the next section, we also present results on this quantity.

The described statistical theory can be used for the optimization of multiplexed single-photon sources, that is, for determining the optimal number of multiplexed units $N_{\opt}$ and the optimal input mean photon number $\lambda_{\opt}$ corresponding to the maximal value of the output single-photon probability $P_{1,\max}$.
The optimum exists because the function $P_1(N,\lambda)$ describing the single-photon probability against the number of multiplexed units and the input mean photon number has a global maximum for most of the systems.
The common characteristics of such systems is that the transmission efficiencies of the various arms change, generally decrease, by increasing the number of PRs in the system. Typical examples are the symmetric and the incomplete multiplexers.
In contrast, for asymmetric (chain-like) multiplexers the same function $P_1(N,\lambda)$ monotonically increases with the number of multiplexed units and it eventually saturates \cite{MazzarellaPRA2013,BonneauNJP2015,AdamPRA2022}.
In such systems $N_\opt$ can be chosen so that the corresponding value of $P_1(N_\opt,\lambda_\opt)$ is reasonably close to the saturated value.

\begin{figure}[tb]
\centering
\includegraphics[width=0.8\textwidth]{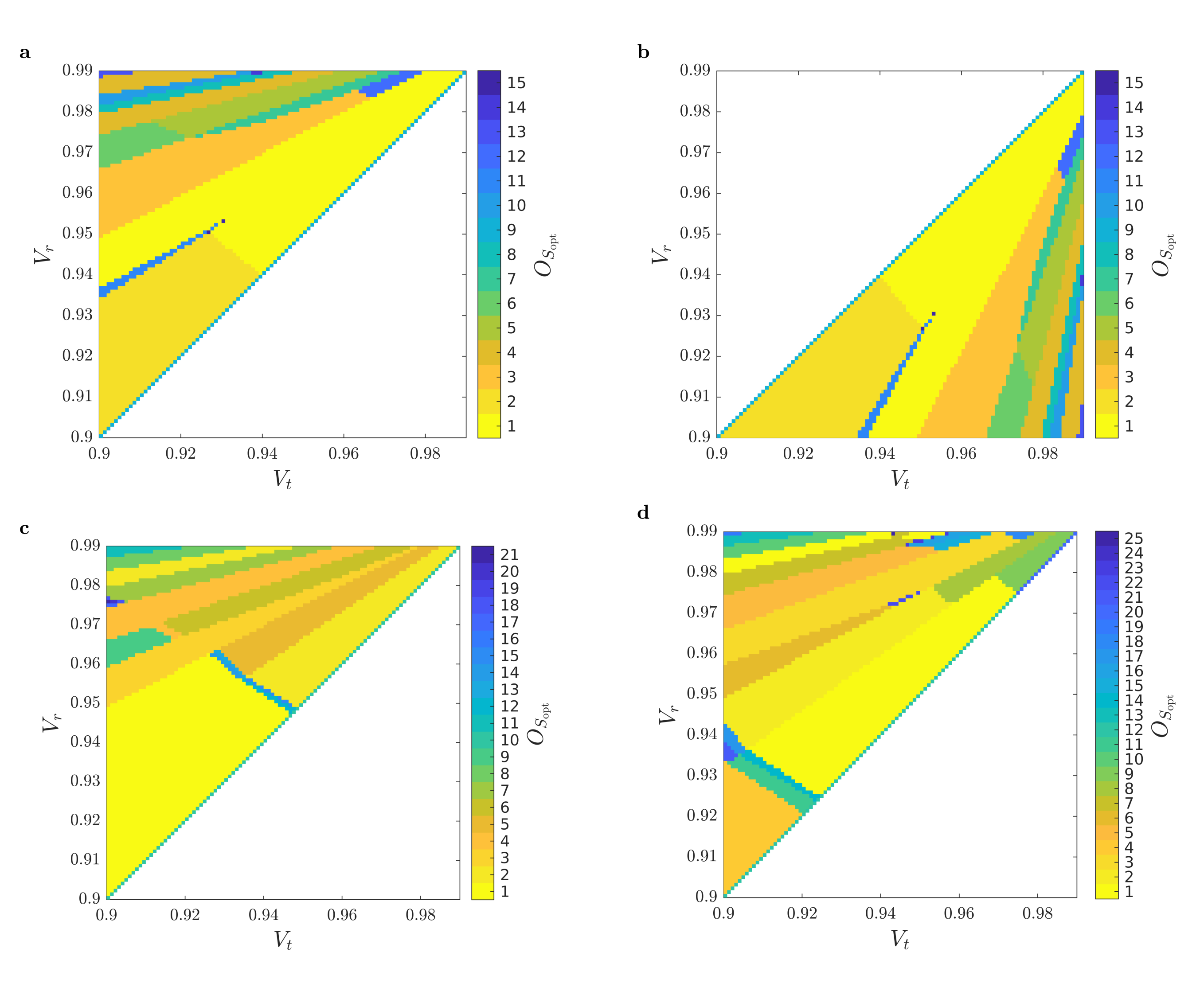}
\caption{Occurrence $O_{S_\opt}$ of the various optimal multiplexer structures $S_\opt$ for SPSs based on OGBMs for (\textbf{a}) the detector efficiency $V_D=0.95$ and $V_r\geq V_t$, (\textbf{b}) $V_D=0.95$ and $V_r\leq V_t$, (\textbf{c}) $V_D=0.85$ and $V_r\geq V_t$, and (\textbf{d}) $V_D=0.8$ and $V_r\geq V_t$, for the number of multiplexed units $N=11$. A particular color denotes a given structure. Increasing sequential numbers of $O_{S_\opt}$ in the color bar represent decreasing occurrence of a specific structure. Color white represents all the structures in the opposite regions. \label{fig:occurregs4}}
\end{figure}
The task for SPSs based on GBMs is to determine the optimal structure for a given number $N_R$ of PRs.
In this paper we determine the optimal structures for fixed values of the number $N_R$ of PRs, that is, for a predefined number $N$ of the MUs.
Hence, we do not address the problem of finding an optimal number of multiplexed units $N_\opt$ for single-photon sources based on GBM.
Note that considering only suboptimal numbers of PRs does not reduce the applicability of our model since typically suboptimal system sizes are realized in current experiments.
We also note that, in accordance with equation~\eqref{eq:Catalan}, the number of geometries strongly increases with the number $N_R$ of the PRs.
For example, for $N_R=15$ the number of possible different geometries is close to ten million.
Hence, finding the optimal number of multiplexed units $N_\opt$ of SPSs based on GBMs can be really cumbersome.

Finding the optimal structure for a given number $N_R$ of PRs can be realized as follows.
First, we generate the sequences representing all GBM structures comprising $N_R$ routers by applying a specific systematic rule.
Based on these sequences, it is possible to calculate the corresponding sets $\{V_n\}$ of total transmission coefficients $V_n$ characterizing the particular multiplexers.
At this point, we mention that some of the sets can contain the same symbolic total transmission coefficients for geometries characterized by different sequences.
As an example, Fig.~\ref{fig:FourExample} shows two different geometries for $N_R=4$ for which the sets $\{V_n\}$ are identical.
In this case, by assuming that the transmission coefficients of the upper and lower inputs of the PRs are denoted by $V_t$ and $V_r$, respectively, the corresponding sets are
a) $\{V_t^2,V_t^2V_r,V_tV_r^2,V_tV_r,V_r^2\}$ and
b) $\{V_t^2,V_tV_r,V_t^2V_r,V_tV_r^2,V_r^2\}$.
Recall that the list of $V_n$s are sorted to a decreasing order before the optimization, therefore SPSs based on multiplexers with identical sets $\{V_n\}$ exhibit the same performance.
Consequently, we consider only the GBM structure appearing first in our logic in the case of identical $\{V_n\}$s.

\begin{figure}[tb]
\centering
\includegraphics[width=\textwidth]{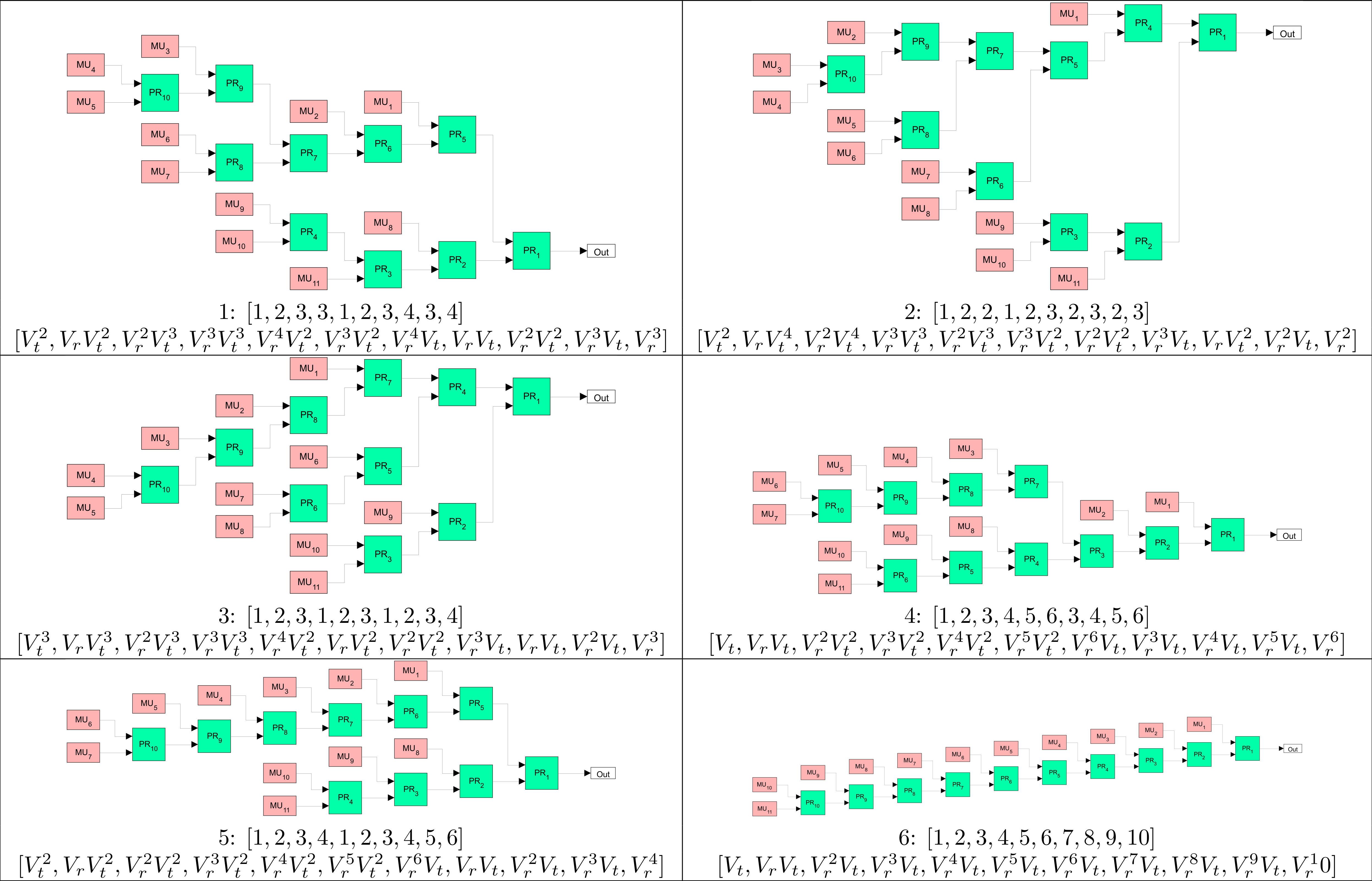}
\caption{The six most frequent optimal structures of OGBMs in the region $V_r\geq V_t$ for the detector efficiency $V_D=0.95$ and the number of multiplexed units $N=11$. $V_t$ and $V_r$ correspond to the upper and lower inputs, respectively, of the individual PRs. The sequential numbers of the occurrence $O_{S_\opt}$ followed by the integer sequences representing the structures, and the lists of the total transmission coefficients $V_n$ in the order of the multiplexed units MU$_n$ are presented below the structures.  \label{fig:tfosVD95upper}}
\end{figure}
After determining the set of total transmission coefficients, we can apply equation~\eqref{general_formula_var} to maximize the single-photon probability $P_{1,S}(\lambda)$ of the SPS based on the given GBM, where we use the subscript $S$ in $P_1$ for denoting the structure.
As in our case the number $N$ of MUs is fixed, the input mean photon number $\lambda$ is the only variable that can be optimized for given values of the transmission coefficients $V_r$ and $V_t$ and the detector efficiency $V_D$.
As the function $P_{1,S}(\lambda)$ has a single maximum, any method for finding extremums can be used to determine the optimal value of $\lambda$.
Finally, after determining the single-photon probability $P_{1,S}(\lambda_\opt)$ that can be achieved for particular $\lambda_\opt$ values for all possible structures $S$, we find the highest one denoted by $P_{1,\max}$. 
The structure $S$ corresponding to this maximal achievable single-photon probability $P_{1,\max}$ is said to be the optimal structure $S_\opt$ of the multiplexer for the given number $N_R$ of PRs or for the corresponding number $N$ of MUs, respectively.
The general binary-tree multiplexer with optimal structure will be termed \emph{optimal general binary-tree multiplexer} and abbreviated as OGBM.
Using this method, one can determine the OGBM for any set of loss parameters characterizing the single-photon source.

\section{Results}

\begin{figure}[tb]
\centering
\includegraphics[width=\textwidth]{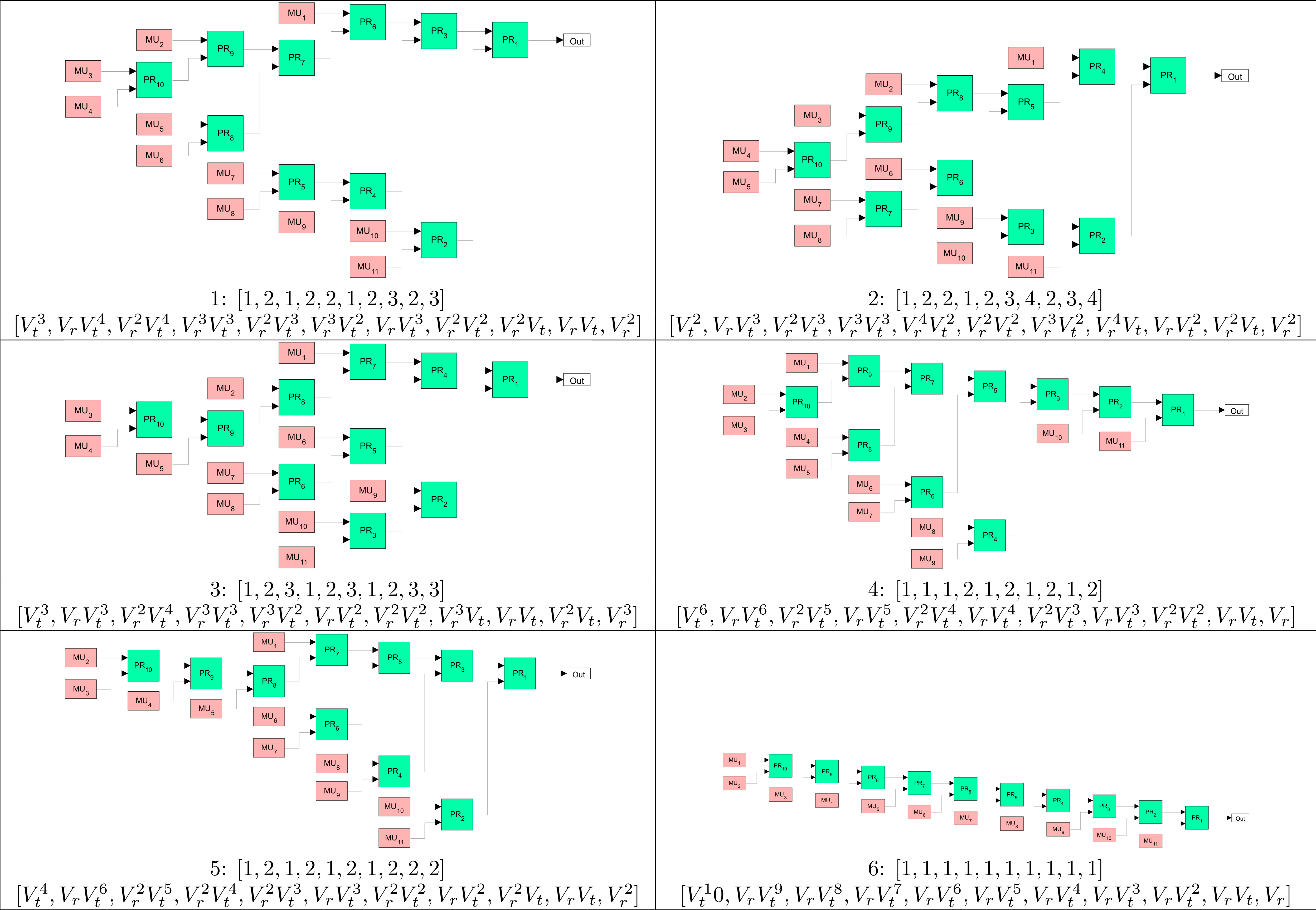}
\caption{The six most frequent optimal structures of OGBMs in the region $V_r\leq V_t$ for the detector efficiency $V_D=0.95$ and the number of multiplexed units $N=11$. $V_t$ and $V_r$ correspond to the upper and lower inputs, respectively, of the individual PRs. The sequential numbers of the occurrence $O_{S_\opt}$ followed by the integer sequences representing the structures, and the lists of the total transmission coefficients $V_n$ in the order of the multiplexed units MU$_n$ are presented below the structures. \label{fig:tfosVD95lower}}
\end{figure}
In this section, we present our results on the optimization of SPSs based on GBMs composed of general asymmetric routers.
Our goal is to find the GBM formed by $N_R$ photon routers that has the best performance, therefore we confine our calculations to high transmission and detector efficiencies that can be realized experimentally with state-of-the-art devices.
Hence, in this section the detector efficiency is set to $V_D=0.95$, the highest value reported in ref.\cite{Fukuda2011}.
However, we apply different values $V_D$ in certain cases we consider as relevant. 
The general transmission coefficient is set to $V_b=0.98$ in all our calculations, hence generally we do not indicate this value in the following.
Routers built of bulk-optical elements exhibited the highest transmission efficiencies $V_r=0.99$ and $V_t=0.985$ reported in refs.\cite{Peters2006, Kaneda2019}
These values are applied in our analysis whenever individual parameter sets or sweeps for other parameters are analyzed.
However, we use $V_r=V_t=0.99$ as the upper boundaries of the ranges of these efficiencies whenever we present parameter sweeps for them to show the symmetry of these parameters, while the lower boundaries of these ranges are chosen to be $V_r=V_t=0.9$ ensuring that single-photon detection yields the highest single-photon probability for the whole considered parameter range (see, e.g., ref.\cite{AdamPRA2022}).

\begin{figure}
\centerline{\includegraphics[width=\textwidth]{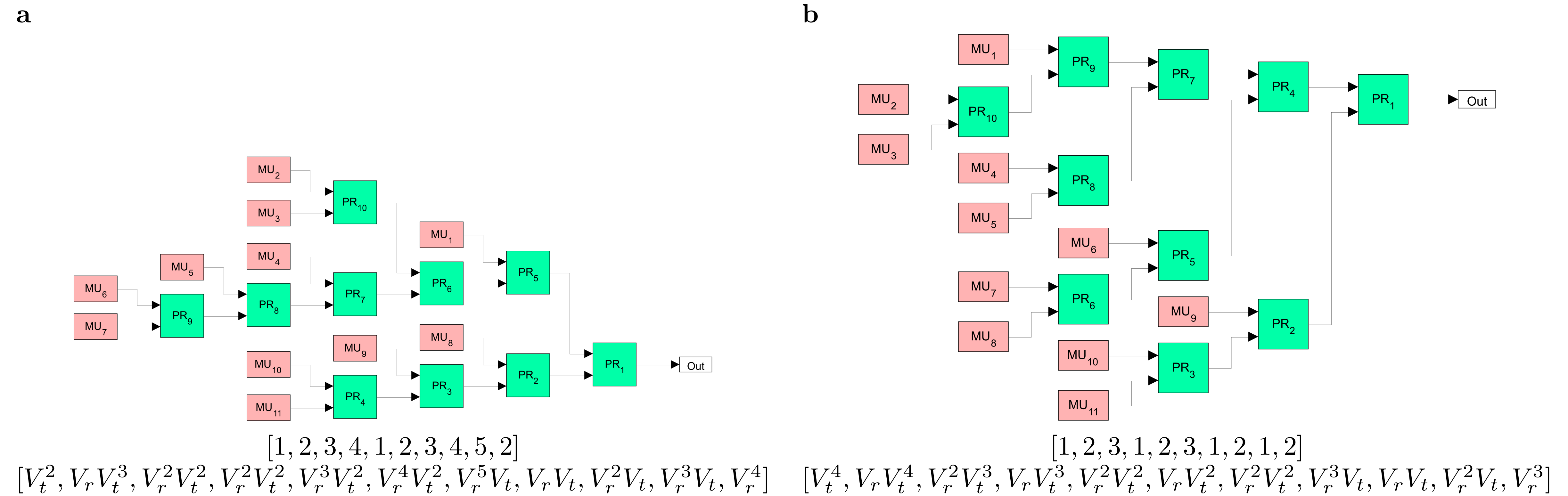}}
\caption{Optimal structures of OGBMs for the detector efficiencies (\textbf{a}) $V_D=0.8$ and (\textbf{b}) $V_D=0.85$, the transmission coefficients $V_r=0.99$ and $V_t=0.985$, and the number of multiplexed units $N=11$. The integer sequences representing the structures and the lists of the total transmission coefficients $V_n$ in the order of the multiplexed units MU$_n$ are presented below the structures. \label{fig:2structexamples}}
\end{figure}
Figs.~\ref{fig:P1g2DP1Dg2}a and \ref{fig:P1g2DP1Dg2}b present the maximal single-photon probability $P_{1,\max}^{\rm ogbm}$ and the corresponding normalized second-order autocorrelation function $g^{(2)}_{\rm ogbm}$, respectively, for SPSs based on OGBMs, while Figs.~\ref{fig:P1g2DP1Dg2}c and \ref{fig:P1g2DP1Dg2}d show the difference $\Delta_P^{\rm ogbm - asym}$ between the maximal single-photon probabilities and the difference $\Delta_{g^{(2)}}^{\rm asym - ogbm}$ between the normalized second-order autocorrelation functions, respectively, for SPSs based on OGBM and ASYM multiplexers as functions of the transmission coefficients $V_t$ and $V_r$ for the detector efficiency $V_D=0.95$, and the number of multiplexed units $N=11$.
It can be seen that, as it is expected, the maximal single-photon probability $P_{1,\max}$ is higher while the second-order autocorrelation function $g^{(2)}$ is better, that is, lower for higher values of the transmission coefficients of the PRs.
The highest maximal single-photon probabilities in this region are above $P_{1,\max}>0.86$ and the lowest corresponding values of the second-order autocorrelation function are $g^{(2)}<0.1$.
Figure~\ref{fig:P1g2DP1Dg2}c shows that using SPSs based on OGBMs give higher single-photon probabilities than SPSs based on ASYM multiplexers for the whole considered parameter range.
From Fig.~\ref{fig:P1g2DP1Dg2}d one can deduce that the normalized second order autocorrelation function values $g^{(2)}$ are smaller for SPSs based on OGBMs than for SPSs based on ASYM multiplexers except for very asymmetric PRs, that is, for $V_r\gg V_t$ or $V_r\ll V_t$. 
The $g^{(2)}$ values can be lower for SPSs based on ASYM multiplexers than for those based on OGBMs because the optimization was carried out for the single-photon probability $P_1$.
We have also compared our results with output-extended incomplete binary-tree multiplexers and symmetric multiplexers for $N=4$ and $N=8$, and we have found that the advantage of SPSs based on OGBMs is on the same range as in the case of ASYM multiplexers.

Next, we show our results on the optimal structures yielding the maximal single-photon probabilities. 
Figure~\ref{fig:occurregs4} presents the occurrence $O_{S_\opt}$ of the various optimal multiplexer structures $S_\opt$ for SPSs based on OGBMs for the detector efficiency $V_D=0.95$ and $V_r\geq V_t$ (Fig.~\ref{fig:occurregs4}a), $V_D=0.95$ and $V_r\leq V_t$ (Fig.~\ref{fig:occurregs4}b), $V_D=0.85$ and $V_r\geq V_t$ (Fig.~\ref{fig:occurregs4}c), and $V_D=0.8$ and $V_r\geq V_t$ (Fig.~\ref{fig:occurregs4}d), for the number of multiplexed units $N=11$.
A particular color denotes a given structure.
Increasing sequential numbers of $O_{S_\opt}$ in the color bar represent decreasing occurrence of a specific structure.
Color white represents all the structures in the opposite regions.
The sizes and shapes of the regions in Figs.~\ref{fig:occurregs4}a and \ref{fig:occurregs4}b are identical but mirrored to the $V_t=V_r$ line (they are reflected congruent shapes) as it is expected from symmetry consideration.
However, the identified structures belonging to a particular color in the two regions can be different due to the fact that the method described in Section~\ref{sec:2} selects a single GBM structure out of those having identical sets $\{V_n\}$ of total transmission coefficients.
This can be deduced from Figs.~\ref{fig:tfosVD95upper} and \ref{fig:tfosVD95lower} where the six most frequent optimal structures of OGBMs occurring in Figs.~\ref{fig:occurregs4}a (region $V_r\leq V_t$) and \ref{fig:occurregs4}b (region $V_r\geq V_t$) are presented, respectively, for the detector efficiency $V_D = 0.95$ and the number of multiplexed units $N = 11$.
In these figures the transmission coefficients $V_t$ and $V_r$ correspond to the upper and lower inputs, respectively, of the individual PRs.
The figures also contain the sequential numbers of the occurrence $O_{S_\opt}$ of the structures shown in Figs.~\ref{fig:occurregs4}a and \ref{fig:occurregs4}b, the integer sequences representing the structure, and the lists of the total transmission coefficients $V_i$ in the order of the multiplexed units MU$_n$.
Note that the sets of total transmission coefficients $\{V_n\}$ in Fig.~\ref{fig:tfosVD95upper} are the same as the corresponding sets in Fig.~\ref{fig:tfosVD95lower} if the roles of the transmission coefficients $V_r$ and $V_t$ are swapped.
This property reflects the expected symmetry mentioned before.
Figs.~\ref{fig:tfosVD95upper} and \ref{fig:tfosVD95lower} show that the asymmetric multiplexer, that is, a chain-like structure, proves to be the best for a certain region of the transmission coefficients $V_t$ and $V_r$.
Obviously, in this region the difference $\Delta_P^{\rm ogbm - asym}$ presented in Fig.~\ref{fig:P1g2DP1Dg2}c is zero.
From Figs.~\ref{fig:occurregs4}c and \ref{fig:occurregs4}d one can deduce that by decreasing the value of the detector efficiency $V_D$ the number of optimal multiplexer structures occurring in the analyzed domain increases, and, obviously, the regions representing particular structures are considerably different.
As an example, in Fig.~\ref{fig:2structexamples} we show the optimal structures of OGBMs for the detector efficiencies $V_D=0.8$ (Fig.~\ref{fig:2structexamples}a) and $V_D=0.85$ (Fig.~\ref{fig:2structexamples}b), the transmission coefficients $V_r=0.99$ and $V_t=0.985$, and the number of multiplexed units $N=11$.
The integer sequences representing the structures and the lists of the total transmission coefficients $V_n$ in the order of the multiplexed units MU$_n$ are presented below the structures.
These structures are apparently different, and they do not occur in Figs.~\ref{fig:tfosVD95upper} or \ref{fig:tfosVD95lower} either.
We have compared the number of structures that are unique in a chosen region for the detector efficiencies $V_D=[0.8,0.85]$ and $V_D=0.95$.
We have found that the number of unique structures present for $V_D=0.8$ ($V_D=0.85$) but absent for $V_D=0.95$ is 18 (11), while there are 8 (5) unique structures present for $V_D=0.95$ and absent for $V_D=0.8$ ($V_D=0.85$).

\begin{figure}
\centering
\includegraphics[width=\textwidth]{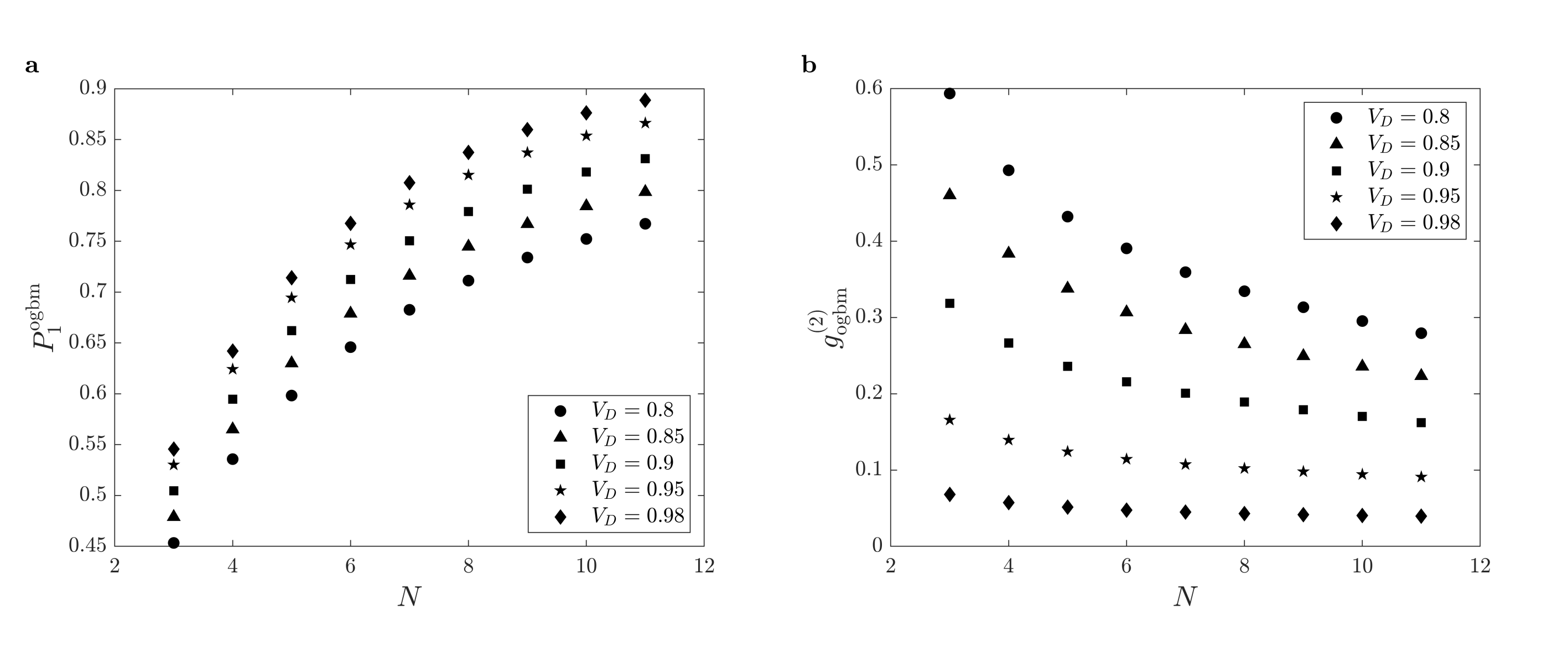}
\caption{(\textbf{a}) The maximal single-photon probabilities $P_{1,\max}^{\rm ogbm}$ and (\textbf{b}) the normalized second-order autocorrelation function $g^{(2)}_{\rm ogbm}$ for SPSs based on OGBMs as functions of the number of multiplexed units $N$ for the transmission coefficients $V_t=0.985$ and $V_r=0.99$, and for various values of the detector efficiency $V_D$. \label{fig:vonalabrak}}
\end{figure}
Finally, in Fig.~\ref{fig:vonalabrak} we show the maximal single-photon probabilities $P_{1,\max}^{\rm ogbm}$ (Fig.~\ref{fig:vonalabrak}a) and the normalized second-order autocorrelation function $g^{(2)}_{\rm ogbm}$ (Fig.~\ref{fig:vonalabrak}b) for SPSs based on OGBMs as functions of the number of multiplexed units $N$ for the transmission coefficients $V_t=0.985$ and $V_r=0.99$, and for various values of the detector efficiency $V_D$.
As it is expected, increasing the number of multiplexed units $N$ leads to increasing values of the maximal single-photon probability $P_{1,\max}^{\rm ogbm}$ and decreasing values of the second order autocorrelation function $g^{(2)}_{\rm ogbm}$.
Also, increasing the detector efficiency $V_D$ relevantly enhances the performance of the SPS.
The maximal single-photon probability $P_{1,\max}$ that can be achieved for the detector efficiency $V_D=0.95$ and the number of multiplexed units $N=11$ is $P_{1,\max}=0.866$, while by modifying the detector efficiency to $V_D=0.98$ this probability can reach $P_{1,\max}=0.889$.
These single-photon probabilities that can be achieved with experimentally realizable system sizes are quite promising compared to the probabilities that can be achieved with completely optimized SPSs based on previously considered multiplexers for the same loss parameters.
For example, assuming the transmission coefficients $V_t=0.985$ and $V_r=0.99$, and the detector efficiency $V_D=0.95$, and using optimized system sizes in SPSs based on ASYM multiplexers, the maximal single-photon probability is $P_{1,\max}^{\rm asym}=0.905$.
However, the corresponding optimal number of the multiplexed units is considerably higher, $N_\opt^{\rm asym}=28$.
The value of the second-order autocorrelation function $g^{(2)}$ that can be achieved for the detector efficiency $V_D=0.95$ and the number of multiplexed units $N=11$ is $g^{(2)}=0.091$, while by modifying the detector efficiency to $V_D=0.98$ this value is $g^{(2)}=0.0395$.
Note that these values can be also promising as they occur at the high single-photon probabilities mentioned above and using a multiplexer of small size.

\section{Conclusion}
To improve the performance of spatially multiplexed single-photon sources, we have developed a method for optimizing the structure of general binary-tree multiplexers realized with asymmetric photon routers.
Our procedure systematically considers all possible binary-tree multiplexers that can be formed by a certain number of photon routers.
Subsequent to optimizing the input mean photon numbers for each of the single-photon sources containing the considered multiplexers and single-photon detectors, one can select the multiplexer structure that leads to the highest single-photon probability for a given set of loss parameters characterizing the system.
We have determined the optimal general binary-tree multiplexers for experimentally realizable values of the transmission coefficients of the photon routers and that of the detector efficiency.
As it is expected, SPSs based on OGBM yield higher single-photon probabilities compared to what can be achieved with SPSs based on any other multiplexer considered in the literature.
Our approach improves the performance of multiplexed single-photon sources even for small system sizes which is the typical situation in current experiments.

\section*{Acknowledgments}
This research was supported by the National Research, Development and Innovation Office, Hungary (``Frontline'' Research Excellence Programme Grant No.\ KKP133827, and Projects No. TKP 2021-NVA-04, TKP2021-EGA-17).

\end{document}